\documentclass[a4paper,10pt,twoside]{cpc-hepnp}
\usepackage{CJK,upgreek,fancyhdr}
\usepackage{multicol}
\usepackage{graphicx}
\usepackage{booktabs}
\usepackage{amssymb,bm,mathrsfs,bbm,amscd}
\usepackage[tbtags]{amsmath}
\usepackage{lastpage}

\begin{document}
\begin{CJK*}{GBK}{song}

\fancyhead[c]{\small Chinese Physics C~~~Vol. xx, No. x (201x) xxxxxx}
\fancyfoot[C]{\small 010201-\thepage}

\footnotetext[0]{Received 14 october 2015}

\title{Evolution of $N = 28$ shell closure in relativistic continuum Hartree-Bogoliubov theory\thanks{Supported by the Major State 973 Program of China (Grant No. 2013CB834400), the National Natural Science Foundation
of China (Grants No. 11175002, No. 11335002, No. 11375015 and No. 11461141002) and Research Fund for the Doctoral Program of Higher Education (20110001110087).}}

\author{%
      Xue-Wei XIA$^{1:1)}$\email{xuewei.xia@buaa.edu.cn}
}
\maketitle

\address{%
$^1$ School of Physics and Nuclear Energy Engineering,
Beihang University, Beijing 100191, China\\

}

\begin{abstract}
The $N = 28$ shell gap in sulfur, argon, calcium and titanium isotopes is investigated in the framework of relativistic continuum Hartree-Bogoliubov (RCHB) theory. The evolutions of neutron shell gap, separation energy, single particle energy and pairing energy are analyzed, and it is found that $N = 28$ shell gap is quenched in sulfur isotopes but persists in argon, calcium and titanium isotopes. The evolution of $N = 28$ shell gap in $N = 28$ isotonic chain is discussed, and the erosion of $N = 28$ shell gap is understood with the evolution of potential with proton number.
\end{abstract}

\begin{keyword}
$N = 28 $ shell closure, RCHB theory, shell evolution
\end{keyword}

\begin{pacs}
21.10.-k, 21.60.Jz, 27.40.+z
\end{pacs}

\footnotetext[0]{\hspace*{-3mm}\raisebox{0.3ex}{$\scriptstyle\copyright$}2013
Chinese Physical Society and the Institute of High Energy Physics
of the Chinese Academy of Sciences and the Institute
of Modern Physics of the Chinese Academy of Sciences and IOP Publishing Ltd}%

\begin{multicols}{2}

\section{Introduction}

Since Goeppert-Mayer~\cite{Mayer1949PhysRev.75.1969} and Haxel et al~\cite{Haxel1949PhysRev.75.1766.2} independently interpreted {\it magic numbers} above 20 with introducing the spin-orbit (SO) coupling in the one-body nuclear potential, shell structure has been considered as an essential aspect of atomic nucleus and the magic numbers have become the cornerstones for the development of nuclear physics. With the development of radioactive ion beam facilities around the world, it creates the possibility for exploring the new shell structure and examining the traditional magic numbers for exotic nuclei far from the $\beta$-stability line~\cite{Tanihata1985PhysRevLett.55.2676}. It is found that away from stability, the traditional magic numbers could disappear, such as $N = 8$~\cite{Navin2000PhysRevLett.85.266,Iwasaki2000PLB}, 20~\cite{GuillemaudMueller1984NPA, Motobayashi1995PLB}, and 28~\cite{Sorlin1993PhysRevC.47.2941, Glasmacher1997PhysicsLettersB163, Sarazin2000PhysRevLett.84.5062, Bastin2007PhysRevLett.99.022503}, and new magic numbers could appear, such as $N = 16$~\cite{Ozawa2000PhysRevLett.84.5493} and 34~\cite{Steppenbeck2013Nature}. Besides, new magic number~\cite{Meng1998PLB,Zhang2005NPA} and new symmetry are topics in nuclear physics~\cite{Ginocchio1997PhysRevLett.78.436, Meng1998PhysRevC.58.R628, Meng1999PhysRevC.59.154, Ginocchio2004PhysRevC.69.034303, Long2006PLB, Liang2011PhysRevC.83.041301, Guo2012PhysRevC.85.021302, Lu2012PhysRevLett.109.072501, Liang2013PhysRevC.87.014334, Shen2013PhysRevC.88.024311, Guo2014PhysRevLett.112.062502, Liang2015report}.

The neutron-rich $N \simeq 28$ nuclei play a crucial role in the nucleosynthesis of the heavy Ca-Ti-Cr isotopes and are now experimentally accessible in the modern radioactive ion beam facilities~\cite{Sorlin1993PhysRevC.47.2941}. A lot of experimental efforts have been devoted to determining whether the $N = 28$ shell closure is eroded in neutron-rich nuclei and if yes, in which specific isotopic chains it is.
Mass measurement and spectroscopic experiments have shown that $N = 28$ shell closure is absent in silicon~\cite{Bastin2007PhysRevLett.99.022503, Campbell2006PhysRevLett.97.112501,Takeuchi2012PhysRevLett.109.182501,Stroberg2014PhysRevC.90.034301} and sulfur isotopes~\cite{Glasmacher1997PhysicsLettersB163, Sarazin2000PhysRevLett.84.5062, Gaudefroy2009PhysRevLett.102.092501, Force2010PhysRevLett.105.102501, Santiago-Gonzalez2011PhysRevC.83.061305}. For argon isotopes, the persisted $N = 28$ shell closure is shown in $\beta$ decay~\cite{Grevy2003NPA}, $\gamma$-ray spectroscopy ~\cite{Bhattacharyya2008PhysRevLett.101.032501,Gade2009PhysRevLett.102.182502}, Coulomb excitation~\cite{Scheit1996PhysRevLett.77.3967,Glasmacher1997PhysicsLettersB163}, knockout~\cite{Gade2005PhysRevC.71.051301} and transfer~\cite{Gaudefroy2006PhysRevLett.97.092501,Gaudefroy2008PhysRevC.78.034307} reaction experiments. However, a life measurement of $2^+_1$ state of $^{46}$Ar deduced a high $B(E2)$ value~\cite{Mengoni2010PhysRevC.82.024308}, which is at odd with the above experiments. Very recently, the mass of $^{47-48}$Ar has been measured, which provides strong evidence for the closed shell nature of neutron number $N = 28$ in argon and demonstrates argon is the lowest even-$Z$ element exhibiting the $N = 28$ shell closure~\cite{Meisel2015PhysRevLett.114.022501}.

Theoretically, a lot of approaches have been employed to describe properties of neutron-rich $N \simeq 28 $ nuclei within the framework of shell model~\cite{Retamosa1997PhysRevC.55.1266, Dean1999PhysRevC.59.2474, Caurier2002EJPA, Caurier2004NPA, Caurier2005RevModPhys.77.427, Nowacki2009PhysRevC.79.014310, Gaudefroy2010PhysRevC.81.064329, Caurier2014PhysRevC.90.014302} and self-consistent mean field theory~\cite{Werner1996NPA, Hirata1996NPA, Lalazissis1999PhysRevC.60.014310, Carlson2000PhysRevC.62.054310, Peru2000EJPA, Rodriguez-Guzman2002PhysRevC.65.024304, Moreno-Torres2010PhysRevC.81.064327, Rodriguez2011PhysRevC.84.051307, LiZP2011PhysRevC.84.054304, Wang2011PhysRevC.83.054305, Wang2014CPL}. In relativistic mean field (RMF) approach, the axial RMF theory suggested $N = 28$ shell closure is broken in $^{42}$Si and $^{44}$S~\cite{Werner1996NPA}. The triaxial RMF theory predicted a shape evolution from prolate shapes for $^{42-44}$S to triaxial shapes for $^{48-52}$S, then an oblate shape for $^{54}$S and to a spherical shape for $^{56}$S~\cite{Hirata1996NPA}. Taking the pairing correlation into account, relativistic Hartree-Bogoliubov (RHB) theory based on harmonic oscillator basis predicted a strong suppression of the $N = 28$ shell gap for neutron-rich magnesium, silicon and sulfur isotopes~\cite{Lalazissis1999PhysRevC.60.014310}. Furthermore, the erosion of the $N = 28$ shell closure and low-energy collective spectra in neutron-rich nuclei are described with the beyond mean field method generator coordinate method~\cite{Rodriguez-Guzman2002PhysRevC.65.024304, Rodriguez2011PhysRevC.84.051307}, which are based on Gogny interaction and relativistic density function based five dimensional collective Hamiltonian~\cite{LiZP2011PhysRevC.84.054304}.

In order to describe the neutron-rich exotic nuclei with $N \simeq 28$, one should consider carefully the coupling of continuum which is crucial for exotic nuclei. Coordinate space, in which the continuum is discretized by suitably large box boundary conditions, is a proper representation for treating continuum~\cite{Dobaczewski1984NPA, Dobaczewski1996PhysRevC.53.2809}.
As an extension of the relativistic mean field and the Bogoliubov transformation in the coordinate representation, relativistic continuum Hartree-Bogoliubov (RCHB) theory provides a self-consistent treatment of pairing correlations in the presence of the continuum.
The RCHB theory has achieved great success in describing the ground-state properties of nuclei far from the $\beta$-stability line, including the first microscopic self-consistent description of halo in $^{11}$Li~\cite{Meng1996PhysRevLett.77.3963}, prediction of giant halos in light and medium-heavy nuclei~\cite{Meng1998PhysRevLett.80.460,Meng1998PhysRevC.65.041302} as well as new magic numbers in superheavy nuclei~\cite{Zhang2005NPA}, and reproduction of interaction cross section and charge-changing cross sections in light exotic nuclei~\cite{Meng1998PLB,Meng2002PLB}. Later, as a generalization to the deformed nuclei, the deformed relativistic Hartree-Bogoliubov theory in continuum (DRHBc) has been developed~\cite{Zhou2010PhysRevC.82.011301,Li2012PhysRevC.85.024312,Chen2012PhysRevC.85.067301}.

In this paper, we will investigate  $N = 28$ shell closure in sulfur, argon, calcium and titanium isotopes with the RCHB theory. The evolution of shell gap will be analyzed and the comparison between theory and experiment will be included. In Section 2, we give the formalism of the RCHB theory. The numerical details are presented in Section 3 and we discuss results for sulfur, argon, calcium and titanium isotopes in Section 4. A summary is given in Section 5.

\section{Theoretical framework}

We briefly describe the framework of RCHB theory. All the details on the RCHB theory can be found in Refs~\cite{Meng1996PhysRevLett.77.3963,Meng1998NPA}. The starting point is a point-coupling Lagrangian density where nucleons are described as Dirac spinors,
\begin{eqnarray}\label{EQ:LAG}
  {\cal L} &=&{\cal L}^{\rm{free}}+{\cal L}^{\rm{4f}}+{\cal L}^{\rm{hot}}+{\cal L}^{\rm{der}}+{\cal L}^{\rm{em}},
\end{eqnarray}
with
 \begin{eqnarray}
{\cal L}^{\rm{free}}&=&\bar\psi(i\gamma_\mu\partial^\mu-M)\psi\\
{\cal L}^{\rm{4f}~~} &=&-\frac{1}{2}\alpha_S(\bar\psi\psi)(\bar\psi\psi)
-\frac{1}{2}\alpha_V(\bar\psi\gamma_\mu\psi)(\bar\psi\gamma^\mu\psi)\nonumber\\
& &
-\frac{1}{2}\alpha_{TV}(\bar\psi\vec{\tau}\gamma_\mu\psi)(\bar\psi\vec{\tau}\gamma^\mu\psi)\\
{\cal L}^{\rm{hot}}&=& -\frac{1}{3}\beta_S(\bar\psi\psi)^3-\frac{1}{4}\gamma_V[(\bar\psi\gamma_\mu\psi)(\bar\psi\gamma^\mu\psi)]^2\nonumber\\
& &-\frac{1}{4}\gamma_S(\bar\psi\psi)^4
\end{eqnarray}
 \begin{eqnarray}
   {\cal L}^{\rm{der}}&=&-\frac{1}{2}\delta_S\partial_\nu(\bar\psi\psi)\partial^\nu(\bar\psi\psi)
-\frac{1}{2}\delta_V\partial_\nu(\bar\psi\gamma_\mu\psi)\partial^\nu(\bar\psi\gamma^\mu\psi)\nonumber\\
& &
-\frac{1}{2}\delta_{TV}\partial_\nu(\bar\psi\vec\tau\gamma_\mu\psi)\partial^\nu(\bar\psi\vec\tau\gamma_\mu\psi)\\
  {\cal L}^{\rm{em}}&=&-\frac{1}{4}F^{\mu\nu}F_{\mu\nu}-e\frac{1-\tau_3}{2}\bar\psi\gamma^\mu\psi A_\mu,
 \end{eqnarray}
where $M$ is the nucleon mass, and $\alpha_S$, $\alpha_V$, $\alpha_{TV}$, $\alpha_{TS}$, $\beta_{S}$, $\gamma_S$, $\gamma_{V}$, $\delta_{S}$, $\delta_{V}$, $\delta_{TV}$, $\delta_{TS}$ are the coupling constants. $A_{\mu}$ and  $F_{\mu\nu}$ are respectively the four-vector potential and field strength tensor of the
electromagnetic field.

Starting from the Lagrangian density~(\ref{EQ:LAG}), one can derive the relativistic Hartree-Bogoliubov (RHB) equation for the nucleons~\cite{KucharekandRing1991ZPA},
\begin{eqnarray}
 \int d^3 \mathbf{r}'
 \left(
  \begin{array}{cc}
   h_D
   - \lambda &
   \Delta
   \\
  -\Delta^*
   & -h_D
   + \lambda \\
  \end{array}
 \right)
 \left(
  { U_{k}
  \atop V_{k}
   }
 \right)
 & = &
 E_{k}
  \left(
   { U_{k}
   \atop V_{k}
    }
  \right),~~~
 \label{eq:RHB0}
\end{eqnarray}
where $E_{k}$ is the quasiparticle energy, $\lambda$ is the chemical potential, and $h_D$ is the Dirac Hamiltonian,
\begin{equation}
 h_D(\mathbf{r}) =
  \bm{\alpha} \cdot \mathbf{p} + V(\mathbf{r}) + \beta (M + S(\mathbf{r})),
\label{eq:Dirac0}
\end{equation}
and the scalar and vector potentials are, respectively,

\begin{eqnarray}
  S(\mathbf{r})&=&\alpha_S \rho_S + \beta_S \rho_S^2 +\gamma_S\rho_S^3 +\delta_S\triangle \rho_S,\\
  \label{eq:vaspot}
V(\mathbf{r})&=&\alpha_V \rho_V + \gamma_V\rho_V^3 +\delta_V \triangle\rho_V+e A_0\nonumber\\
           & &+\alpha_{TV}\tau_3 \rho_{TV}+\delta_{TV} \tau_3 \triangle \rho_{TV},
   \label{eq:vavpot}
\end{eqnarray}
with the local densities
  \begin{eqnarray}\nonumber
      \rho_S(\mathbf{r})     &=&\sum_{k>0 }\bar V_k(\mathbf{r})V_k(\mathbf{r}),\\
      \rho_{V}(\mathbf{r})   &=&\sum_{k>0 } V_k^{\dagger}(\mathbf{r})V_k(\mathbf{r}),\\ \nonumber
      \rho_{T V}(\mathbf{r}) &=&\sum_{k>0 } V_k^{\dagger}(\mathbf{r})\tau_3 V_k(\mathbf{r}),
      \label{eq:mesonsource}
  \end{eqnarray}
where, according to the no-sea approximation, the sum over $k > 0$ runs over the quasiparticle states corresponding to single particle energies in and above the Fermi sea.

The pairing potential reads
\begin{eqnarray}
\Delta_{kk'}(\mathbf{r},\mathbf{r'})
&=&-\sum_{\tilde{k}\tilde{k'}}\mathbf{V}_{kk',\tilde{k}\tilde{k'}}(\mathbf{r},\mathbf{r'})\kappa_{\tilde{k}\tilde{k'}}(\mathbf{r},\mathbf{r'}),
\end{eqnarray}
with the pairing tensor $ \kappa=U^{*}V^T$ and a density-dependent delta pairing force
\begin{eqnarray}
    V^{pp}(\mathbf{r_1},\mathbf{r_2})&=&V_0\delta(\mathbf{r_1}-\mathbf{r_2})\frac{1}{4}(1-P^{\sigma})
                                        (1-\frac{\rho(\mathbf{r_1})}{\rho_0}).~~~
\end{eqnarray}

Considering spherical symmetry, the quasiparticle wave function can be written as
\begin{eqnarray}
\psi_U^i=\left(
\begin{array}{c}
i\frac{G_U^{ijl}(r)}{r}Y_{jm}^l(\theta,\phi)\\
\frac{F_U^{ijl}(r)}{r}(\bm{\sigma\cdot \hat{r}})Y_{jm}^l(\theta,\phi)
\end{array}\right)\chi_t(t), \nonumber\\
\psi_V^i=\left(
\begin{array}{c}
i\frac{G_V^{ijl}(r)}{r}Y_{jm}^l(\theta,\phi)\\
\frac{F_V^{ijl}(r)}{r}(\bm{\sigma\cdot \hat{r}})Y_{jm}^l(\theta,\phi)
\end{array}\right)\chi_t(t).
\end{eqnarray}
Then, the RHB equation depends only on the radial coordinates and can be expressed as the following integral-differential equations:
\begin{eqnarray}\nonumber
\frac{dG_U}{dr}+\frac{\kappa}{r}G_U(r)-(E+\lambda-V(\mathbf{r})+S(\mathbf{r}))F_U(r)\nonumber\\
+r\int r'dr'\Delta_F(\mathbf{r},\mathbf{r'})F_V(r')=0 \nonumber\\
\frac{dF_U}{dr}-\frac{\kappa}{r}F_U(r)+(E+\lambda-V(\mathbf{r})-S(\mathbf{r}))G_U(r)\nonumber\\
+r\int r'dr'\Delta_G(\mathbf{r},\mathbf{r'})G_V(r')=0 \nonumber\\
\frac{dG_V}{r}+\frac{\kappa}{r}G_V (r)+(E-\lambda+V(\mathbf{r})-S(\mathbf{r}))F_V (r)\nonumber\\
+r\int r'dr'\Delta_F(\mathbf{r},\mathbf{r'})F_U(r')=0\nonumber \\
\frac{dF_V}{r}-\frac{\kappa}{r}F_V (r)-(E-\lambda+V(\mathbf{r})+S(\mathbf{r}))G_V (r)\nonumber\\
+r\int r'dr'\Delta_G(\mathbf{r},\mathbf{r'})G_U(r')=0.
\end{eqnarray}
If the zero-range pairing force is used, the above coupled integral-differential equations are reduced to differential ones, which can be directly solved in coordinate space.

\section{Numerical details}
In order to describe the continuum and its coupling to the bound states properly, the solution of RCHB equation (15) is carried out in coordinate space within a spherical box of radius $R = 20$ fm and a step size of 0.1 fm. It is checked that for these values of the grid, suitable convergence is achieved for all the results presented here. In the present calculation, for particle-hole channel the density functional PC-PK1~\cite{Zhao2010PhysRevC.82.054319} is used, which has achieved great success in describing not only nuclear ground-state but also various of excited-state
properties, including nuclear mass~\cite{Zhao2012PhysRevC.86.064324, Hua2012scichia, Qu2013scichia, Zhang2014Frontier, Lu2015PhysRevC.91.027304, Afanasjev2015PhysRevC.91.014324}, nuclear low-lying excited states~\cite{Yao2013PLB, Yao2014PhysRevC.89.054306, Wu2014PhysRevC.89.017304,Li2012PLB,Fu2013PhysRevC.87.054305,Li2013PLB,Xiang2013PhysRevC.88.057301,Wang2015JPG},  quadrupole and magnetic moments~\cite{Zhao2014PhysRevC.89.011301,LiJ2013PhysRevC.88.064307}, magnetic~\cite{Zhao2011PLB, Steppenbeck2012PhysRevC.85.044316, Yu2012PhysRevC.85.024318, LiJ2013PhysRevC.88.014317} and antimagnetic~\cite{Zhao2011PhysRevLett.107.122501, Zhao2012PhysRevC.85.054310, Li2012PhysRevC.86.057305, Zhang2014PhysRevC.89.047302, Peng2015PhysRevC.91.044329} rotation bands, multiple chiral doublet bands~\cite{Meng2006PhysRevC.73.037303, Kuti2014PhysRevLett.113.032501}, stapler band~\cite{Chen2015PhysRevC.91.044303}, pairing transition at finite temperature~\cite{Niuyf2013PhysRevC.88.034308}, exotic shape~\cite{ZhaoJ2012PhysRevC.86.057304}, fission barrier~\cite{Lu2012PhysRevC.85.011301, Lu2014PhysRevC.89.014323, ZhaoJ2015PhysRevC.91.014321}, shape coexistence and $\alpha$ decay~\cite{Zhang2013PhysRevC.88.054324, Lizx2015Frontier}, isoscalar proton-neutron pairing and $\beta$ decay~\cite{Liang2008PhysRevLett.101.122502, Liang2009PhysRevC.79.064316, Niuzm2013PhysRevC.87.051303, Song2014PhysRevC.90.054309, Yao215PhysRevC.91.024316}, etc. For particle-particle channel, the density-dependent delta pairing force (13) is used with the saturation density $\rho_{0}$ equals to empirical value 0.152 fm$^{-3}$. The pairing strength $V_0 = 685.0$ MeV fm$^3$ is fixed by reproducing the experimental odd-even mass differences of Ca isotopes, Sn isotopes, $N = 20$ isotones and $N = 50$ isotones.

\section{Results and Discussion}
Systematical calculation for sulfur, argon, calcium and titanium isotopes is performed using RCHB theory with effective interaction PC-PK1~\cite{Zhao2010PhysRevC.82.054319}. The calculated two-neutron separation energy $S_{2n}$ and the neutron shell gap defined by one-neutron separation energy $S_n$, $D_n(Z,N)=(-1)^{N+1}[S_{n}(Z,N+1)-S_n(Z,N)]$~\cite{Brown2013PhysRevLett.111.162502}, are in good agreement with available experimental values.
\begin{center}
\includegraphics[width=0.35\paperwidth]{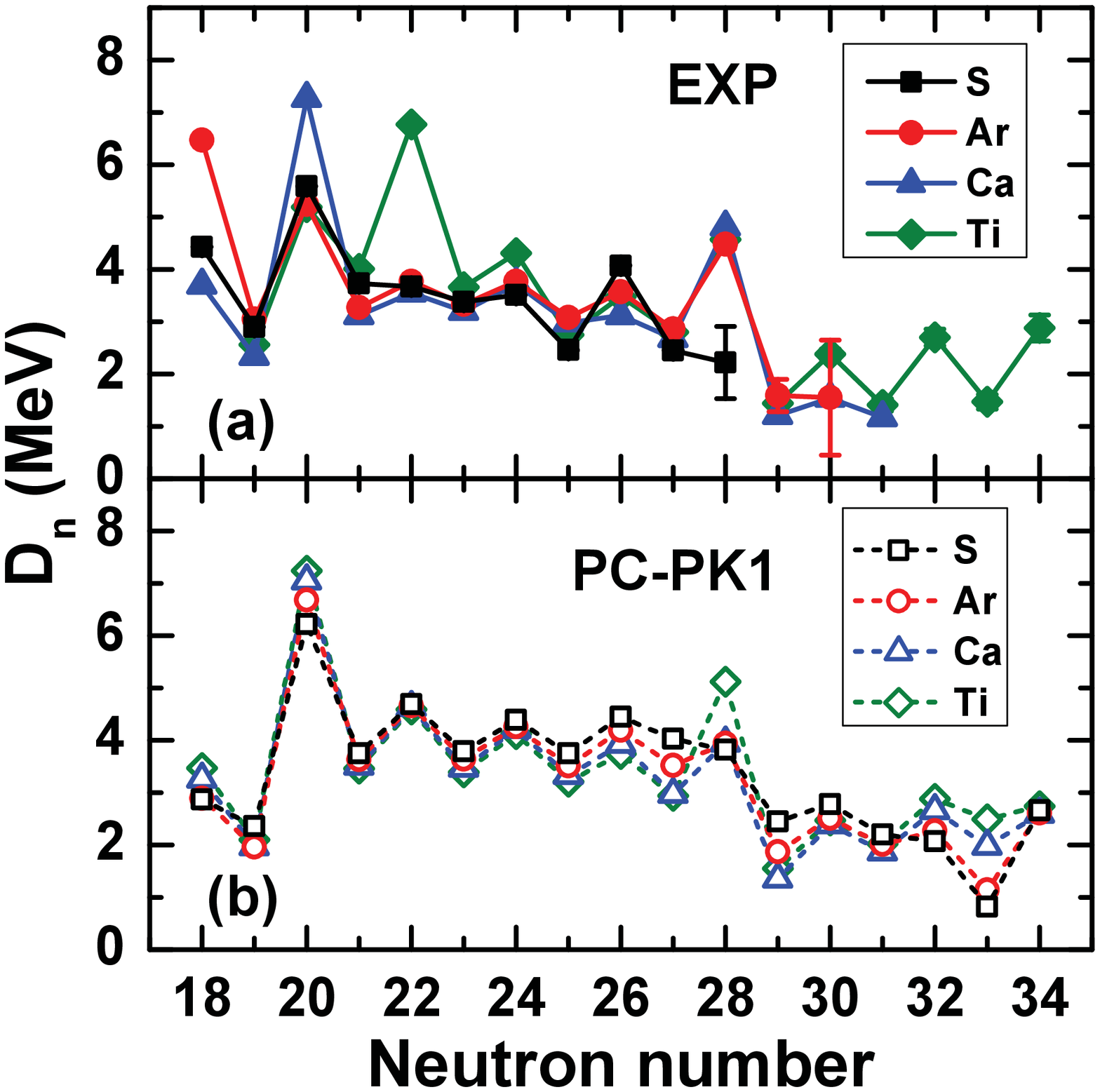}
\figcaption{\label{fig1} (Color online) The evolution of neutron shell gap $D_n=(-1)^{N+1}[S_{n}(Z,N+1)-S_n(Z,N)]$ as a function of neutron number for sulfur, argon, calcium, and titanium obtained in RCHB calculation with effective interaction PC-PK1, and $D_n$ from experimental data is also shown for comparison.}
\end{center}

Neutron shell gap $D_n$ provides a readily recognizable signature of a shell closure~\cite{Brown2013PhysRevLett.111.162502}. A peak in $D_n$ at a certain neutron number along with a change in the $D_n$ level before and after that neutron number gives a good indication of a shell gap~\cite{Brown2013PhysRevLett.111.162502}.

Fig.~\ref{fig1} shows the values of $D_n$ for sulfur, argon, calcium and titanium isotopes obtained from experiment and RCHB calculation with effective interaction PC-PK1. The experimental data are from AME2012 mass table~\cite{Wang2012CPC} and recent data for $^{48-49}$Ar~\cite{Meisel2015PhysRevLett.114.022501}. As shown in Fig.~\ref{fig1}(a) and discussed in Ref.~\cite{Meisel2015PhysRevLett.114.022501}, $D_n$ peaks at $N = 28$ together with a significant reduction of its value after $N = 28$ for argon, calcium and titanium isotopes, which clearly indicates the $N = 28$ neutron shell closure in these isotopes, while the vanish of these signatures for sulfur isotopes demonstrates sulfur does not exhibit the $N = 28$ closed shell. $D_n$ from RCHB calculation presents the same evolution trend with the experimental results, $D_n$ for argon, calcium and titanium from RCHB theory reaches a peak at $N = 28$ and follows a drop for $N > 28$, and sulfur isotope does not present these features. One may examine the $N = 20$ shell closure in these isotopic chains as well.
As it also can be seen in Fig.~\ref{fig1}, both the experimental and theoretical results show the peaks at $N = 20$ for sulfur, argon, calcium and titanium isotopes, which indicates the existence of $N=20$ shell gap in these isotopic chains. It is noted that, the experimental value of $D_n$ for calcium at $N = 20$ is larger than the other three isotopes, while the large experimental value of $D_n$ is also found for argon and titanium at $N = 18$ and $N=22$, respectively. But these behaviors do not exist in theoretical results, the reason may be the absence of Wigner term~\cite{Wigner1937PhysRev.51.106, Isacker1995PhysRevLett.74.4607, Goriely2002PhysRevC.66.024326} in the calculations.

Two-neutron separation energy $S_{2n}(Z,N)=BE(Z,N)-BE(Z,N-2)$ is also a widely used probe of neutron shell gap. Generally, the trend of $S_{2n}$ with neutron number $N$ is smooth, except after having passed a major shell gap. There, when adding two more neutrons, the $S_{2n}$ value drops significantly. A remarkable decline of $S_{2n}$ indicates the occurrence of a neutron shell closure.
\begin{center}
\includegraphics[width=0.35\paperwidth]{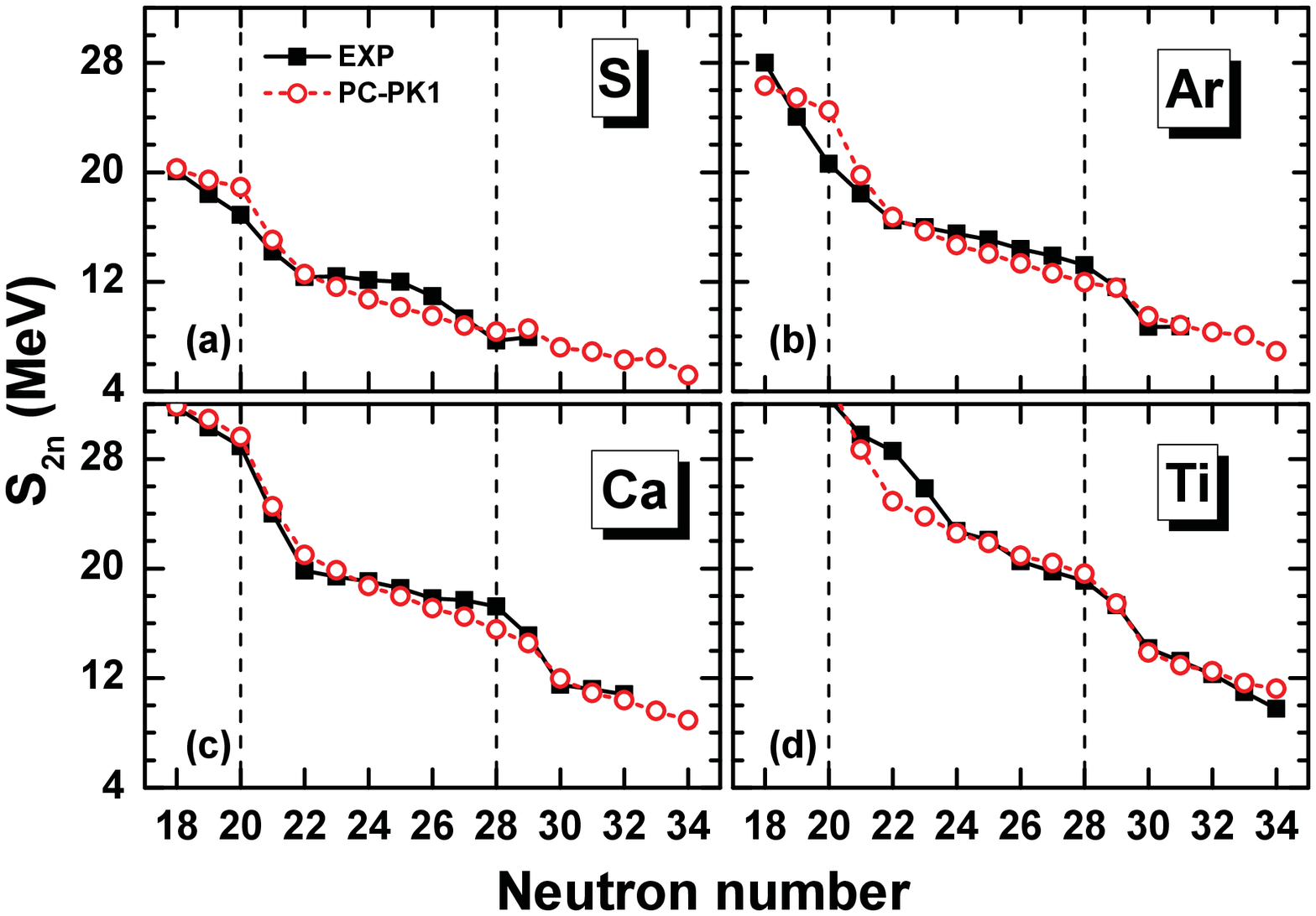}
\figcaption{\label{fig2} (Color online) The two-neutron separation energy $S_{2n}$ trend  for sulfur, argon, calcium and titanium from RCHB calculation employing effective interaction PC-PK1 (open circles) is shown along with $S_{2n}$ from experimental masses(solid squares).}
\end{center}

Fig.~\ref{fig2} shows the two-neutron separation energy $S_{2n}$ for sulfur, argon, calcium and titanium isotopes from RCHB theory and experimental data~\cite{Wang2012CPC}. As it can be seen that good agreement between theory and experiment exists for sulfur, argon, calcium and titanium isotopes. A sharp drop in $S_{2n}$ from both RCHB calculation and experiment occurs for argon, calcium and titanium isotopic chains when crossing $N=28$, which reveals the $N=28$ closed shell. On the other hand, the evolution of $S_{2n}$ for sulfur isotopic chain displays a different pattern. An increase in $S_{2n}$ is observed in the sulfur chain when crossing $N=28$ both for theory and experiment, which clearly shows the vanish of $N=28$ shell closure in sulfur isotopes.

In Fig.~\ref{fig3}, the calculated neutron single particle states for $^{40-50}$S, $^{42-52}$Ar, $^{44-54}$Ca and $^{46-56}$Ti are shown, as well as the neutron chemical potentials from RCHB theory. It can be found that with the increase of neutron number, chemical potentials for sulfur, argon, calcium and titanium isotopes rise gradually. From the titanium isotope to sulfur isotope with a given neutron number, the chemical potential is getting higher and close to the continuum. For nuclei $^{50}$Ti, the chemical potential $\lambda_n$ coincides with the energy of the last occupied single particle orbit 1f$_{7/2}$ because neutron paring collapse occurs in its RCHB calculation.
\begin{center}
\includegraphics[width=0.35\paperwidth]{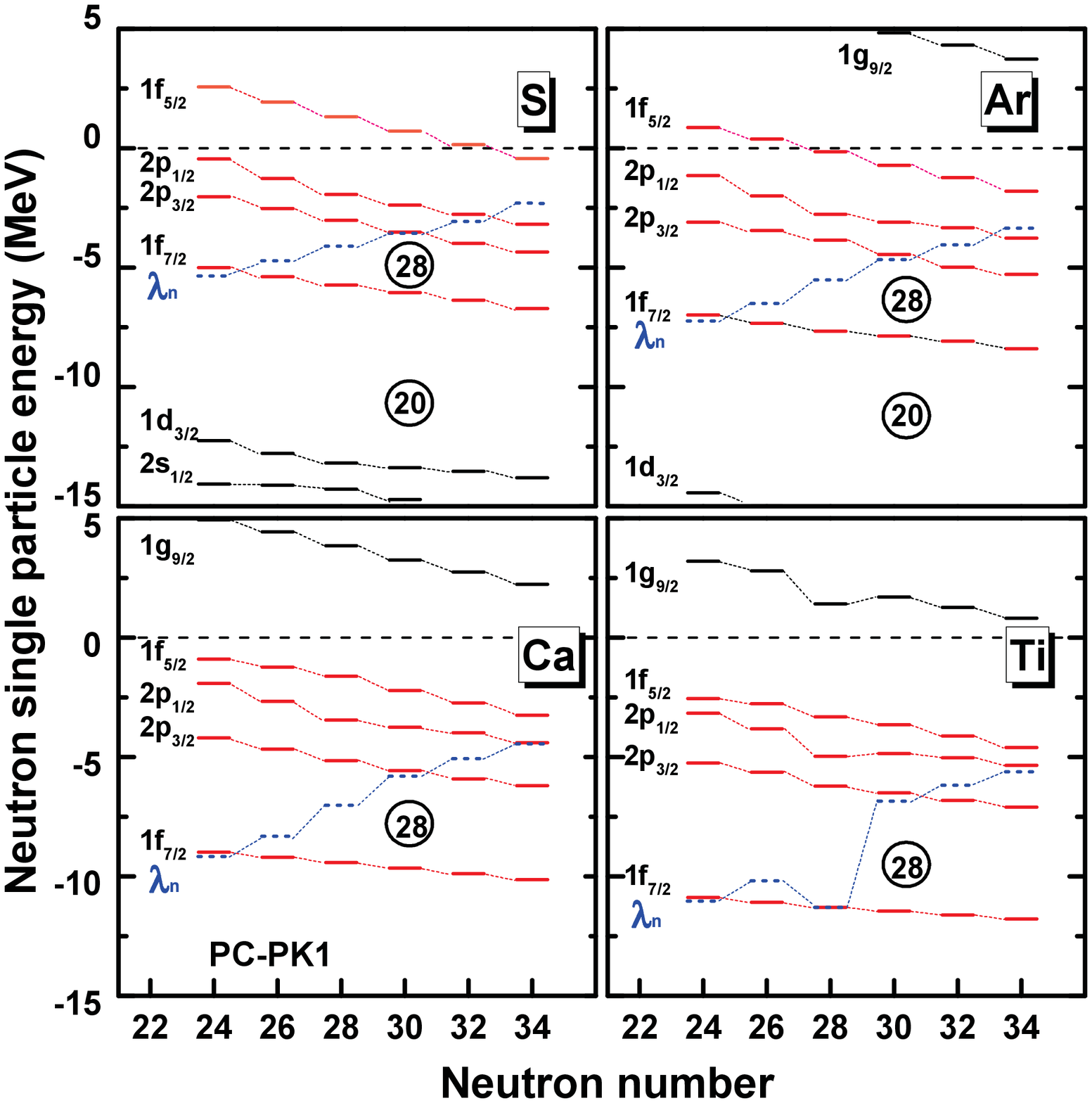}
 \figcaption{\label{fig3}(Color online) Neutron single particle states from RCHB calculation with effective interaction PC-PK1 as the function of neutron number for sulfur, argon, calcium and titanium isotopes. The neutron chemical potential $\lambda_n$ is shown by dashed lines. For nucleus $^{50}$Ti, neutron chemical potential $\lambda_n$ is equal to the energy of last occupied level 1f$_{7/2}$ due to pairing collapse.}
\end{center}

As it can be seen, with the decrease of proton number, single particle levels 1f$_{5/2}$, 2p$_{1/2}$, 2p$_{3/2}$ and 1f$_{7/2}$ are gradually rising and approaching to the continuum, meanwhile the shell gap between 2p$_{3/2}$ and 1f$_{7/2}$ is reduced as the proton number decreases. To give insight into the evolution of shell gap between 2p$_{3/2}$ and 1f$_{7/2}$, an average shell gap for different isotopes is introduced here, which is defined as $\langle\Delta\rangle=(\sum_{N=24}^{N=34}\Delta_N)/N$, where $\Delta_N$ is the shell gap between 2p$_{3/2}$ and 1f$_{7/2}$ for a given nucleus. For titanium isotopes, the average shell gap between 2p$_{3/2}$ and 1f$_{7/2}$ is 5.10 MeV, for calcium and argon isotopes the value are 4.26 and 3.54 MeV, respectively. However, the average shell gap for sulfur is only 2.74 MeV, which indicates that $N = 28$ shell closure is quenched in sulfur isotopes.

Pairing energy which usually vanishes at the closed shell and has a maximum value in the middle of two closed shells, can be used as a probe of magic number as well~\cite{Zhang2005NPA}. In Fig.~\ref{fig4}, the neutron pairing energies for sulfur, argon, calcium and titanium isotopes obtained from RCHB theory with PC-PK1 are shown as the function of neutron number. It can be seen clearly, for titanium, calcium and argon isotopes, the pairing energies peak at $N = 28$ and are close to zero. However, for sulfur isotope the peak at $N = 28$ vanishes and pairing energy at $N = 28$ is very close to the $N = 26$ maximum value, which reflects that sulfur does not exhibit $N = 28$ shell closure.
\begin{center}
\includegraphics[width=0.35\paperwidth]{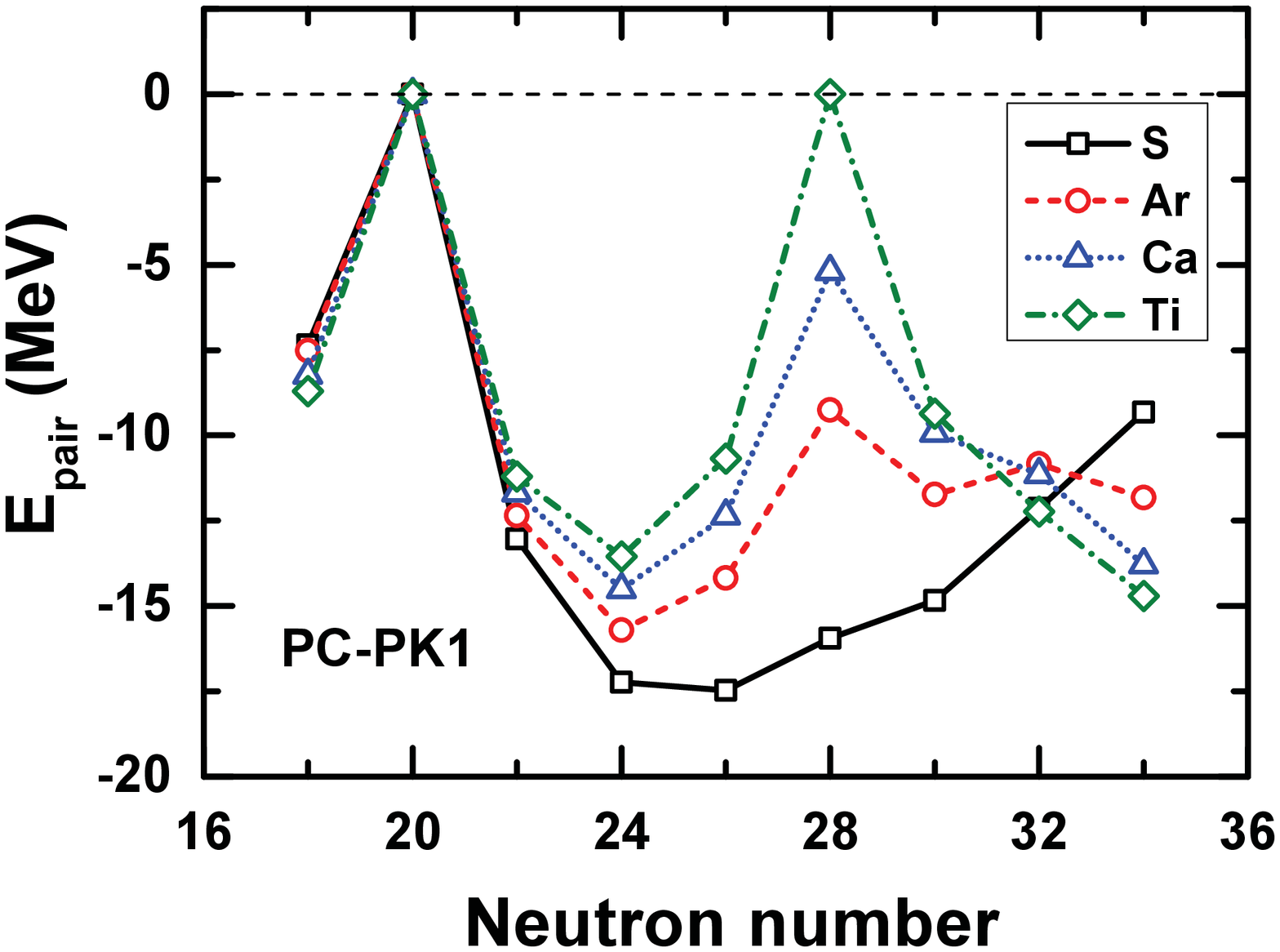}
 \figcaption{\label{fig4}(Color online) Neutron pairing energy in RCHB calculation with effective interaction PC-PK1 as the function of neutron number for sulfur, argon, calcium and titanium isotopes.}
\end{center}

 According to the discussions above, it can be found that, with the decrease of proton number, $N =28$ shell gap decreases dramatically in the RCHB calculation. As mentioned in the introduction, the axial RMF and triaxial RMF theories as well as RHB theory based on harmonic oscillator basis have shown the erosion of $N = 28$ shell closure in sulfur isotope. The present RCHB calculation indicates that the reduction of $N =28$ shell closure in sulfur isotope still exists even though with spherical assumption. Further investigation of deformed relativistic Hartree-Bogoliubov theory in continuum~\cite{Zhou2010PhysRevC.82.011301,Li2012PhysRevC.85.024312,Chen2012PhysRevC.85.067301} is necessary to figure out the effects of deformation.
\begin{center}
\includegraphics[width=0.35\paperwidth]{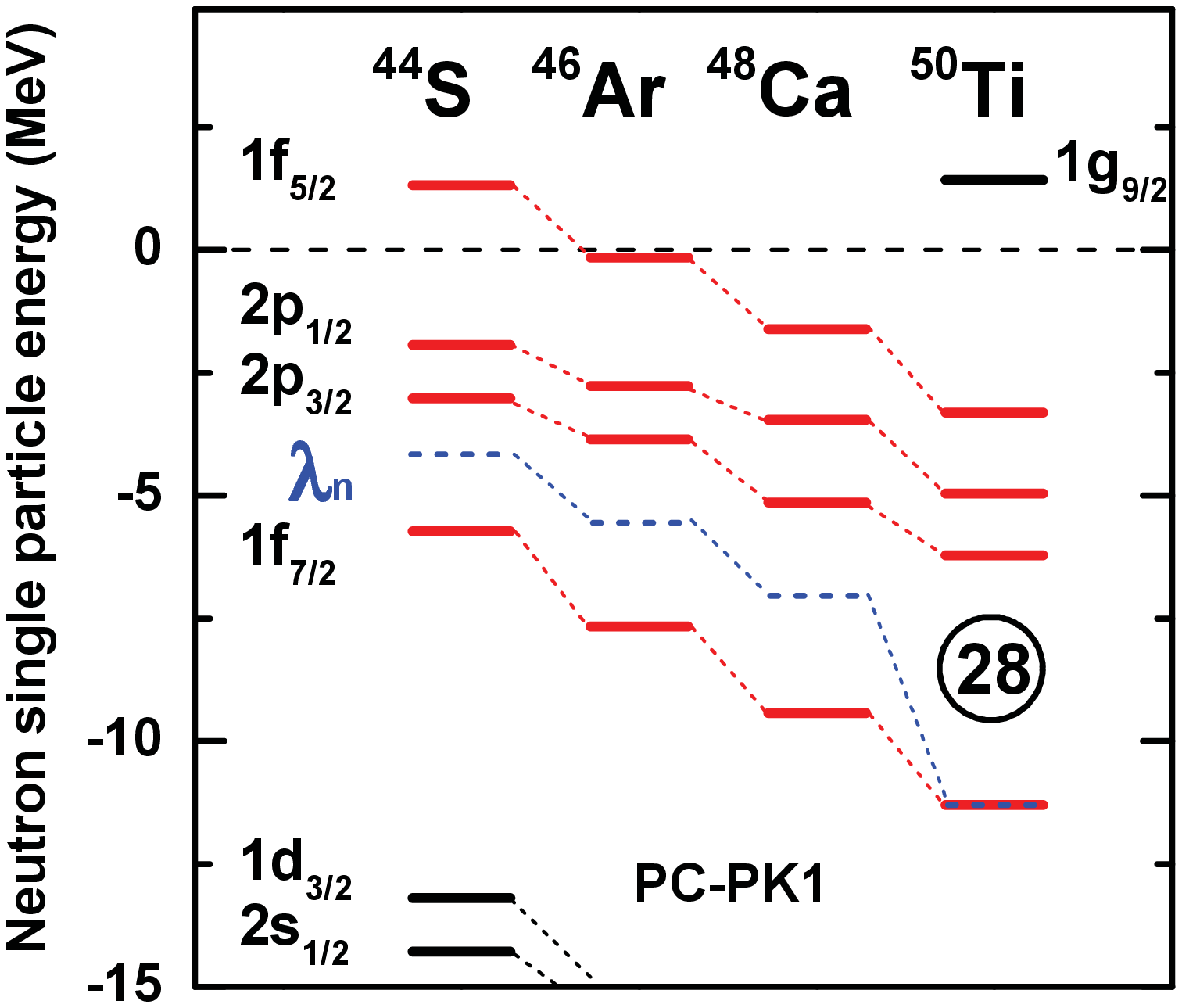}
 \figcaption{\label{fig5}(Color online) Neutron single particle states in nuclei $^{44}$S, $^{46}$Ar, $^{48}$Ca and $^{50}$Ti obtained from RCHB calculation with effective interaction PC-PK1. The neutron chemical potential $\lambda_n$ is shown by dashed lines.}
\end{center}

Furthermore, to investigate the evolution of $N=28$ shell gap more deeply, the neutron single particle orbits of $^{44}$S, $^{46}$Ar, $^{48}$Ca and $^{50}$Ti are shown in Fig.~\ref{fig5}. As it can be seen that, with the decrease of proton number, the shell gap between 2p$_{3/2}$ and 1f$_{7/2}$ decreases rapidly. From $^{50}$Ti to $^{44}$S, the shell gap between 2p$_{3/2}$ and 1f$_{7/2}$ is reduced from 5.10 MeV to 2.71 MeV. It also should be noted that, single particle level 1f$_{7/2}$ is increasing rapidly from $^{50}$Ti to $^{44}$S, but level 2p$_{3/2}$ is going up slightly with the decrease of proton number. As a consequence, the $N = 28$ shell closure is weakening obviously in sulfur isotope.

 The reason for the change of single particle levels 1f$_{7/2}$ and 2p$_{3/2}$ with the decrease of proton number may be traced to the change of potential. By deducing the Dirac equation to the Schr\"odinger-like equation for the upper component, the effective potential felt by single nucleon could be expressed as $V(r)+S(r)+V_{cent}(r)+V_{ls}(r)$~\cite{Meng2006PPNP, ZhangY2010IJMPE}. In Fig.~\ref{fig6}, the effective potentials of 1f$_{7/2}$ and 2p$_{3/2}$ for $^{44}$S, $^{46}$Ar, $^{48}$Ca and $^{50}$Ti are shown, together with the corresponding single particle levels. It can be seen that the effective potentials for 1f$_{7/2}$ and 2p$_{3/2}$ are well separated due to the centrifugal barrier $V_{cent}(r)$ and spin-orbit potential $V_{ls}(r)$ which are relative to orbital angular momentum. The potentials for 1f$_{7/2}$ are higher than those for 2p$_{3/2}$. From $^{50}$Ti to $^{44}$S, the potentials for both 1f$_{7/2}$ and 2p$_{3/2}$ are getting more shallow, and the single particle levels are also rising gradually.
\begin{center}
\includegraphics[width=0.35\paperwidth]{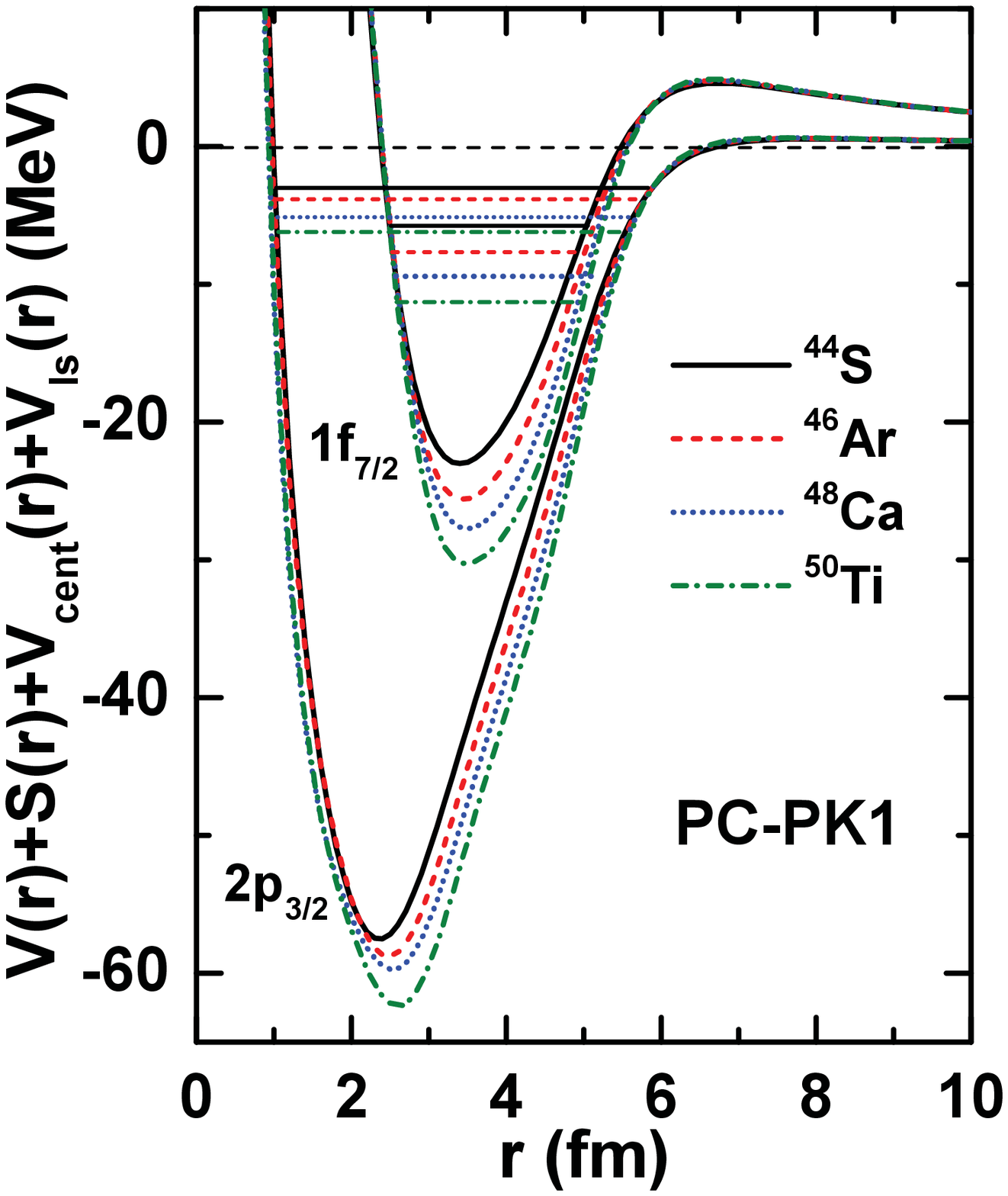}
 \figcaption{\label{fig6}(Color online) Effective potential $V(r)+S(r)+V_{cent}(r)+V_{ls}(r)$ of 1f$_{7/2}$ and 2p$_{3/2}$ for $^{44}$S, $^{46}$Ar, $^{48}$Ca and $^{50}$Ti obtained from RCHB calculation with effective interaction PC-PK1 are shown, together with corresponding single particle levels.}
\end{center}
In order to investigate the evolutions of single particle levels with the change of potentials, the change of single particle energy $\Delta \varepsilon=\varepsilon^{(i)}-\varepsilon^{(^{50}\rm Ti)}$, potential depth $\Delta V_0=V_0^{(i)}-V_0^{(^{50}\rm Ti)}$ and surface diffuseness $\displaystyle\Delta(\frac{r_2}{r_1})=(\frac{r_2}{r_1})^{(i)}-(\frac{r_2}{r_1})^{(^{50}\rm Ti)}$ for 1f$_{7/2}$ and 2p$_{3/2}$ are shown in Fig.~\ref{fig7}, where $i$ represents nuclei $^{44}$S, $^{46}$Ar, $^{48}$Ca and $^{50}$Ti, $r_1$ and $r_2$ correspond to the radial coordinate at 0.1 times potential depth and 0.9 times potential depth, respectively. It can be seen from Fig.~\ref{fig7}(a) that, for 1f$_{7/2}$  $\Delta \varepsilon$ linearly increases with the variation of proton number, and for 2p$_{3/2}$ $\Delta \varepsilon$ also increases but the slope is less than 1f$_{7/2}$. In Fig.~\ref{fig7}(b), the change of potential depth $\Delta V_0$ displays a similar evolution tendency with $\Delta \varepsilon$ for both 1f$_{7/2}$ and 2p$_{3/2}$, which indicates that the change of potential depth may be the major reason for the the change of single particle energy.

A different evolution trend can be found for the change of surface diffuseness $\displaystyle\Delta(\frac{r_2}{r_1})$ in Fig.~\ref{fig7}(c), that from $^{50}$Ti to $^{44}$S, $\displaystyle\Delta(\frac{r_2}{r_1})$ only increases slightly for 1f$_{7/2}$ but increases rapidly for 2p$_{3/2}$. Taking both the potential depth and the surface diffuseness into account, it can be found that the relatively small change of potential depth and relatively large change of surface diffuseness for 2p$_{3/2}$ result in its slower rise in single particle energy than 1f$_{7/2}$. As a consequence, the $N = 28$ shell closure is eroded in sulfur isotope.
\begin{center}
\includegraphics[width=0.35\paperwidth]{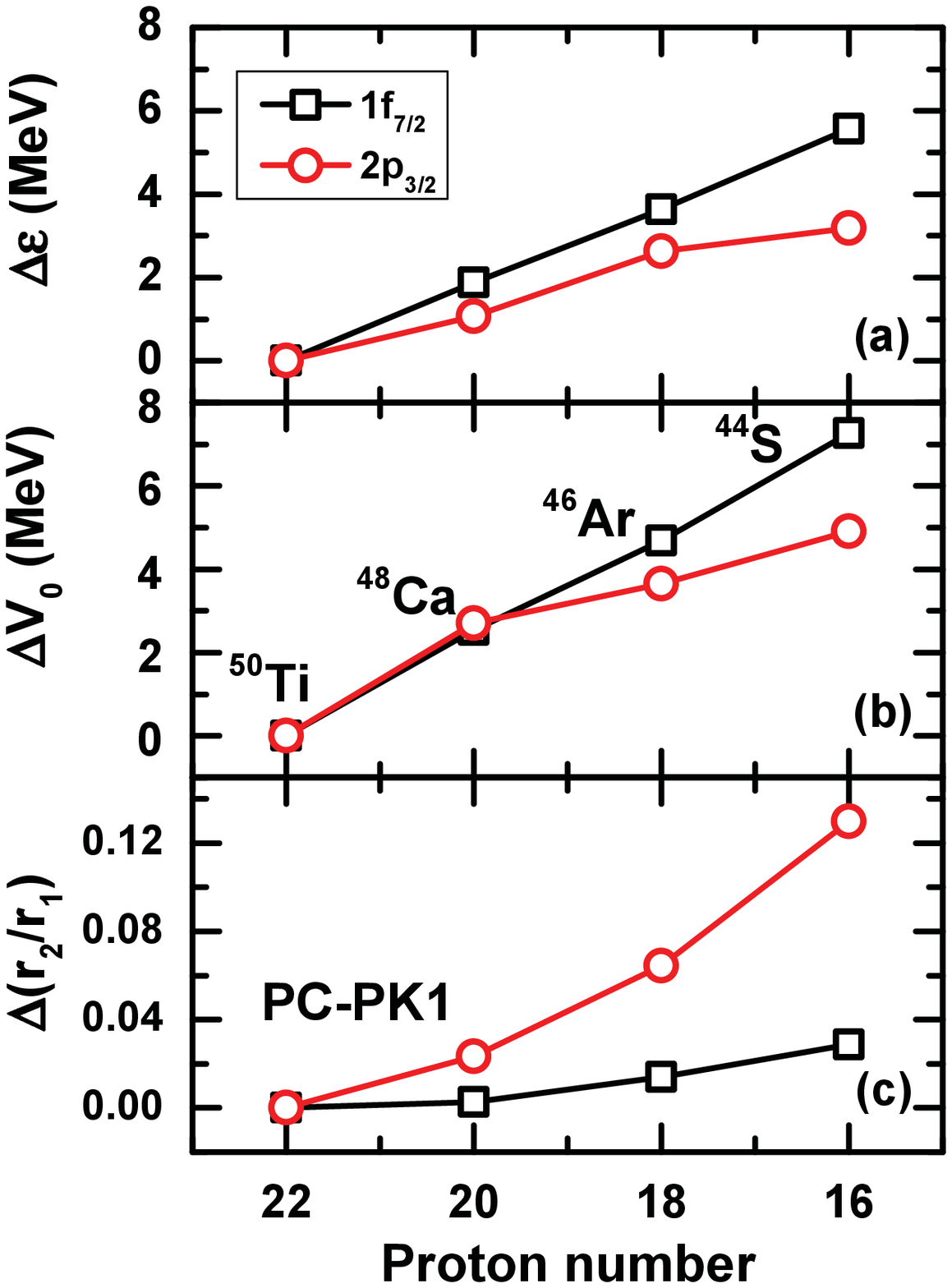}
 \figcaption{\label{fig7}(Color online) The change of single particle energy $\Delta \varepsilon=\varepsilon^{(i)}-\varepsilon^{(^{50}\rm Ti)}$, potential depth $\Delta V_0=V_0^{(i)}-V_0^{(^{50}\rm Ti)}$ and surface diffuseness $\displaystyle\Delta(\frac{r_2}{r_1})=(\frac{r_2}{r_1})^{(i)}-(\frac{r_2}{r_1})^{(^{50}\rm Ti)}$  for 1f$_{7/2}$ and 2p$_{3/2}$ are shown as the function of proton number, where i represents nuclei $^{44}$S, $^{46}$Ar, $^{48}$Ca and $^{50}$Ti, $r_1$ and $r_2$ correspond to the radial coordinate at 0.1 times potential depth and 0.9 times potential depth, respectively.}
\end{center}

\section{Summary}
In summary, the $N = 28$ shell gap in sulfur, argon, calcium and titanium isotopes is investigated with the relativistic continuum Hatree-Bogoliubov (RCHB) theory. Good agreement in the two-neutron separation energy $S_{2n}$ and the neutron shell gap $D_n$  between theory and experiment exists
for sulfur, argon, calcium and titanium isotopes. The evolutions of neutron shell gap, two-neutron separation energy, single particle energy and pairing energy have been analyzed, and a consistent result that $N = 28$ shell gap is eroded in sulfur isotopes while preserves in argon, calcium and titanium isotopes is obtained from the analysis. The evolution of $N = 28$ shell gap in $N = 28$ isotonic chain is also discussed. As the proton number decreases, the relatively small change of potential depth and relatively large change of surface diffuseness for 2p$_{3/2}$ result in its slower rise in single particle energy than 1f$_{7/2}$, then leads to the erosion of $N = 28$ shell closure. It should be noted that, the results here indicate from $^{50}$Ti to $^{44}$S, with the decrease of proton number the $N = 28$ shell closure is quenched, even in the spherical case. Certainly, it is worth to self-consistently consider continuum and deformation effects on the erosion of $N = 28$ shell gap in neutron-rich nuclei, which could be investigated with the DRHBc theory~\cite{Zhou2010PhysRevC.82.011301,Li2012PhysRevC.85.024312,Chen2012PhysRevC.85.067301}.

\acknowledgments{Precious guidance and advice from J. Meng and S. Q. Zhang and helpful discussions with  Z. X. Ren, S. H. Shen and B. Zhao are acknowledged.}

\end{multicols}

\vspace{-1mm}
\centerline{\rule{80mm}{0.1pt}}
\vspace{2mm}

\begin{multicols}{2}

\end{multicols}

\clearpage
\end{CJK*}
\end{document}